\documentclass[12pt]{article}
\usepackage{amsmath,amssymb}
\usepackage{color}

\begin{document}

\title{Quantum Gravity from Fractal Entanglement Geometry}

\author{Jaume Gin\'e}
\date{ Departament de Matem\`atica, Universitat de Lleida, \\
Avda. Jaume II, 69; 25001 Lleida, Catalonia, Spain \\
{\small {\rm E--mail:} {\tt jaume.gine@udl.cat}}}

\maketitle

\begin{abstract}
In this work, we explore the possibility that spacetime geometry may emerge from the entanglement structure of an underlying quantum-information network. We develop an exploratory framework, termed Fractal Entanglement Quantum Gravity (FEQG), in which information-theoretic correlations provide a candidate pregeometric structure from which aspects of geometry, quantum dynamics, and gravity may arise. An information-based distance on the correlation graph is considered as a possible means of relating quantum correlations to geometric separation. The framework also investigates whether scale-dependent and fractal properties of the underlying structure could lead to dimensional flow, potentially approaching an effective dimension of two near the Planck scale. Within a scale-relativistic perspective, nondifferentiable trajectories and stochastic descriptions are examined as possible ingredients in an effective quantum dynamics; however, the emergence of the Schr\"odinger equation from the network structure remains to be established. Similarly, the relation between evolving correlations and gravitational dynamics is explored through candidate generalized field equations, while recovering Einstein gravity in an appropriate macroscopic limit remains an open problem. Potential phenomenological consequences include dimensional reduction, modifications to gravitational potentials, and possible deviations from standard quantum mechanics at ultrashort scales. These possibilities are not presented as established predictions, but as directions for further investigation. FEQG thus provides a research program for examining whether quantum-information structure, fractal geometry, and gravitational phenomena can be connected within a common framework.
\end{abstract}

\section{Introduction}

\subsection{Geometry from Entanglement}

Recent developments in quantum gravity, particularly within the holographic paradigm, have suggested that spacetime geometry may be closely related to underlying patterns of quantum entanglement. This idea is most clearly illustrated by the AdS/CFT correspondence \cite{Maldacena1998}, in which certain geometric properties of a gravitational bulk spacetime are encoded in quantum-information-theoretic quantities of a lower-dimensional boundary theory. A central result in this context is the Ryu--Takayanagi formula \cite{RyuTakayanagi2006},
\begin{equation}\label{RyuTa}
S_A = \frac{\mathrm{Area}(\gamma_A)}{4G},
\end{equation}
which relates the entanglement entropy of a boundary region \(A\) to the area of the corresponding extremal surface \(\gamma_A\) in the bulk.

The significance of this relation lies in the connection it establishes between a quantum-information quantity and a geometric property. The entanglement entropy \(S_A\), defined from the reduced state of region \(A\), characterizes its quantum correlations with the complementary degrees of freedom. In the holographic setting, this information is related to the area of a bulk extremal surface. Although this result does not establish that spacetime geometry can be reconstructed from entanglement in complete generality, it provides a concrete example in which geometric information is encoded in quantum correlations. It therefore motivates the broader possibility that aspects of spacetime geometry may emerge from the informational structure of an underlying quantum state.

The connection extends beyond the original static formulation. Covariant generalizations of the Ryu--Takayanagi prescription, together with studies of perturbations around suitable holographic backgrounds, have revealed further links between entanglement and gravitational dynamics. In particular, the first law of entanglement, $\delta S_A = \delta \langle H_A \rangle$, relates small variations of the entanglement entropy to variations of the expectation value of the modular Hamiltonian \(H_A\), associated with the reduced density matrix \(\rho_A\) of region \(A\). In certain holographic settings, when this relation is combined with the appropriate geometric and dynamical assumptions, it can be used to obtain linearized gravitational field equations. This provides an important connection between quantum-information constraints and the response of bulk geometry, although the result depends on the specific holographic framework and does not imply that arbitrary changes in entanglement automatically generate gravitational dynamics.

The first law of entanglement is formally analogous to the first law of thermodynamics: a small change in the state produces a corresponding change in entropy. Its holographic significance is that, under suitable conditions, this informational relation can be translated into constraints on geometric perturbations. Thus, entanglement is not only related to geometric quantities but may also encode information about how geometry responds to changes in the quantum state.

These results motivate the possibility that a more complete description of spacetime geometry could involve the full pattern of correlations within an underlying quantum system, rather than only the areas of particular extremal surfaces. In this sense, the holographic correspondence provides an important conceptual starting point for investigating whether geometric structures can be reconstructed from quantum-information data \cite{VanRaamsdonk2010}. The extension from surface-area relations to a reconstruction of the full metric, however, remains a nontrivial problem and requires additional assumptions about the quantum degrees of freedom, their correlations, and the emergence of a suitable continuum description.

This question provides one of the motivations for the {\em Fractal Entanglement Quantum Gravity} (FEQG) research program developed in this work. Rather than assuming that the complete spacetime metric is already determined by entanglement, FEQG explores whether an appropriate network of quantum correlations could provide a candidate pregeometric structure from which metric properties may emerge. The precise relation between such informational structures and continuum spacetime geometry is an open issue that will be addressed in the following sections.

\subsection{Quantum and Gravity from Fractal Geometry}

An alternative line of research investigates the possibility that spacetime possesses an underlying fractal or scale-dependent structure, and that quantum behaviour may be related to the non-differentiable geometry associated with such a description \cite{G0,G}. This perspective is closely connected to the theory of scale relativity, in which quantum phenomena are interpreted through the dynamics of geodesics in a spacetime whose geometric properties depend on the resolution scale \cite{Nottale1993,Nottale2011}. Within this approach, quantum mechanics is not introduced simply as an independent modification of classical physics, but is investigated as a possible consequence of a more general geometric framework. Together with the holographic perspective discussed above, these ideas motivate the research program FEQG developed in this work.

The central motivation is that classical differentiability may cease to provide an adequate description of trajectories at sufficiently small scales. In a non-differentiable or fractal setting, the usual limit defining a pathwise velocity need not exist. This motivates the development of generalized kinematical and dynamical descriptions capable of accounting for the irregularity and scale dependence of trajectories. Scale relativity provides one such framework, proposing that quantum behaviour can be understood in terms of motion in a non-differentiable spacetime. It is important to emphasize, however, that non-differentiability alone does not uniquely determine quantum mechanics; additional assumptions concerning the statistical and dynamical properties of the trajectories are required.

A characteristic feature of fractal approaches to spacetime is the possibility of scale-dependent geometric properties. In this context, one may consider an effective metric whose description depends on the probing scale, together with an effective dimension that changes under variations of resolution. This dimensional flow is particularly interesting because several approaches to quantum gravity have reported, under specific definitions and in particular regimes, an effective dimensional reduction towards approximately two dimensions at short distances.

Such ultraviolet dimensional reduction has been investigated in Causal Dynamical Triangulations \cite{Amb,Amb2,Loll}, Asymptotic Safety \cite{Reu,RS}, and Loop Quantum Gravity \cite{Rov,Ash,AL}. These results provide an important motivation for considering scale-dependent geometry within FEQG. Nevertheless, the precise meaning of effective dimension differs among these approaches, and their results do not establish that spacetime must universally become two-dimensional at the Planck scale. Accordingly, the possibility of a similar dimensional flow in FEQG should be regarded as a hypothesis to be investigated rather than an assumed consequence of its fractal structure.

The possible connection between non-differentiable geometry and quantum dynamics provides a further motivation for this framework. In stochastic descriptions of irregular trajectories, fluctuations may exhibit Brownian-type scaling, suggesting that ordinary differential operators could require generalization. Scale-relativity approaches develop this idea through a formulation involving distinct forward and backward descriptions of motion and an associated complex representation of the dynamics. This construction offers a possible route towards quantum equations from geometric and stochastic assumptions, but it is not a universal mathematical consequence of non-differentiability. The relevant stochastic structure and dynamical principles must be specified independently.

This perspective is consistent with earlier proposals relating fractal trajectories to quantum dynamics \cite{Ord1983}. It also connects with broader investigations of quantum gravity based on fractal, anomalous, or multi-scale geometric structures \cite{Calcagni2010}. These approaches provide complementary motivations for examining whether quantum behavior and gravitational phenomena can be understood within a common framework in which geometry depends on scale.

Within FEQG, the central question is whether the informational geometry introduced in the preceding discussion can support such a scale-dependent description. In particular, we investigate the possibility that microscopic irregularities in the emergent geometry may be related to effective quantum dynamics, while its large-scale behavior may admit a smoother geometric description suitable for discussing gravitational phenomena. Establishing this connection requires more than the existence of fractal structure: the microscopic dynamics, the appropriate continuum limit, and the emergence of effective geometric field equations must all be specified.

Thus, fractal geometry and scale relativity provide a complementary perspective to holographic emergent geometry within FEQG. The former motivates the investigation of scale-dependent geometry and its possible relation to quantum dynamics, while the latter motivates the interpretation of geometric structure in terms of quantum correlations. The central hypothesis is that these perspectives may be connected through an underlying informational structure. Whether this connection can consistently account for both quantum dynamics and gravitational phenomena remains an open mathematical and physical problem.

\subsection{A natural bridge between them}

To explore a possible connection between holographic emergent geometry and fractal approaches to spacetime, we propose that the hierarchical organization of quantum correlations may provide a common informational substrate for both perspectives. This proposal is motivated by the holographic insight that, in certain theoretical settings, geometric properties are related to quantum entanglement \cite{RyuTakayanagi2006,VanRaamsdonk2010}. Rather than assuming that entanglement directly produces a smooth geometry, we investigate whether networks of quantum correlations may give rise to an underlying information geometry whose properties depend on the scale at which it is examined.

The motivation for this connection is that both holographic and fractal approaches emphasize the importance of structure across different scales. In holographic theories, geometric quantities can encode information about quantum correlations, while in fractal approaches the effective properties of geometry may change with resolution. These observations suggest that a hierarchical organization of entanglement could provide a setting in which scale-dependent geometric properties emerge. Whether such an organization necessarily leads to fractal geometry, however, remains an open question.

Tensor networks provide a useful mathematical setting in which to investigate this possibility. In particular, the Multiscale Entanglement Renormalization Ansatz (MERA) organizes quantum degrees of freedom through successive coarse-graining layers and represents correlations across multiple length scales \cite{Swingle2012}. Its network structure admits a geometric interpretation that, in appropriate settings, resembles a discretized hyperbolic geometry. MERA therefore illustrates how hierarchical quantum correlations can be associated with an effective network geometry.

Nevertheless, the geometric interpretation of MERA should not be confused with a demonstration that physical spacetime is fractal. Its recursive structure provides a multiscale organization, but self-similarity alone does not establish a non-integer dimension, a particular curvature, or an emergent spacetime metric. These properties depend on the network architecture and on the mathematical criteria used to characterize its geometry.

More generally, tensor networks suggest that geometric properties need not always be introduced independently of the underlying quantum state. Instead, they may be investigated through the organization and connectivity of its correlations. Within FEQG, this motivates the hypothesis that an appropriate quantum-information network could encode an effective notion of spatial proximity and, under suitable conditions, support a scale-dependent geometric description. The precise mathematical construction of this correspondence is developed in the following sections.

A natural starting point for such a construction is the relation between correlation strength and spatial separation. In quantum field theory, correlation functions often exhibit exponential decay with distance in theories with a finite correlation length, at sufficiently large separations. In such regimes, a schematic relation of the form \(I(i,j)\sim e^{-d(i,j)/\ell_0}\) motivates a logarithmic correspondence between distance and correlation strength,
\begin{equation}\label{dist}
d(i,j)\sim-\ell_0\log\left(\frac{I(i,j)}{I_0}\right),
\end{equation}
where \(I(i,j)\) denotes mutual information, \(I_0\) is a reference value, and \(\ell_0\) sets the characteristic length scale. Thus, the logarithmic relation is motivated by the information-theoretic inversion of an exponential correlation decay law: strongly correlated subsystems are associated with smaller effective separations, whereas weakly correlated subsystems are associated with larger ones. This argument provides a heuristic motivation rather than a universal identity, since correlation functions and mutual information may exhibit different scaling behaviors depending on the quantum state and physical regime. Moreover, the resulting expression does not automatically satisfy all the axioms of a metric space, an issue that will be addressed in the subsequent construction.

The conceptual shift is that adjacency need not be defined exclusively through a pre-existing spatial manifold. One may instead investigate whether relational information between quantum subsystems can determine an effective notion of closeness. In this sense, geometry is treated as a candidate emergent structure rather than an initial assumption. Establishing this proposal requires showing that the resulting relational structure admits a consistent geometric interpretation and an appropriate continuum limit.

If the underlying correlation network exhibits nontrivial scaling behavior, its effective dimensionality may differ from that of an ordinary smooth manifold. Such a possibility provides a natural point of contact between holographic emergent geometry and fractal approaches to spacetime. However, the existence and scale dependence of a non-integer dimension must be established through an explicit analysis of the network rather than inferred solely from its hierarchical organization.

A further question is whether changes in the correlation structure can be related to effective gravitational dynamics. This possibility is motivated by entanglement-equilibrium arguments, in which gravitational field equations are connected to constraints on entanglement entropy under specific assumptions \cite{Jacobson1995}. These results provide conceptual motivation for investigating the relation between quantum information and gravity, but they do not establish that arbitrary variations in quantum correlations generate spacetime curvature or reproduce Einstein's equations. Within FEQG, such a dynamical connection remains a conjecture requiring an appropriate geometric and dynamical formulation.

The proposed framework can therefore be understood as an attempt to investigate several related questions: whether quantum correlations can encode geometric connectivity, whether their hierarchical organization can produce scale-dependent or fractal geometric properties, and whether the resulting structure can support an effective description of gravity. The connections between these stages are not assumed to follow automatically; they constitute the central problems that the FEQG program seeks to address.

From this perspective, holographic entanglement geometry and fractal approaches to spacetime may be regarded as complementary sources of ideas for constructing an information-based description of geometry. Their compatibility within a single framework, however, remains to be demonstrated. In particular, the emergence of quantum dynamics, spacetime curvature, and gravitational interactions from the same underlying correlation structure is an open problem rather than an established result.

The possible connection with the ER=EPR conjecture provides an additional motivation for this investigation \cite{MSuss,Suss}. In its original setting, ER=EPR proposes a relation between certain entangled quantum systems and nontraversable Einstein--Rosen bridges. This conjecture suggests a deep connection between quantum correlations and spacetime connectivity, but it does not imply that every nonzero correlation between two subsystems corresponds to a physical wormhole. Nor does it establish that the reduced density matrix of an arbitrary subsystem determines the microscopic topology of spacetime.

Accordingly, within FEQG, ER=EPR is regarded as a conceptual motivation for exploring whether aspects of geometric connectivity can be encoded in quantum correlations. The information network introduced in this work may provide a useful mathematical representation of such relational structure, but identifying its links with microscopic wormholes would require additional geometric and dynamical justification. The broader hypothesis is that the hierarchical organization of quantum correlations may encode geometric information across multiple scales, potentially connecting holographic ideas with scale-dependent and fractal descriptions of spacetime.

\section{The Mathematical Research Program}

The framework developed here is an exploratory program in which geometric structure is investigated as a possible emergent property of quantum correlations, while the role of fractal and scale-dependent features in a potential description of quantum gravity remains to be established. Its starting hypothesis is that spacetime geometry may not be fundamental, but could instead arise from the relational structure of an underlying quantum state. To formulate this possibility, consider a decomposition of a Hilbert space into subsystems,
\begin{equation}
\mathcal{H}=\bigotimes_i \mathcal{H}_i,
\end{equation}
where each \(\mathcal{H}_i\) represents a quantum subsystem, such as a qubit, an oscillator, a field mode, or a spin degree of freedom.

This tensor-product structure provides the mathematical setting in which subsystems and their correlations can be defined. It is important to emphasize, however, that such a decomposition is not generally unique or determined by the Hilbert space alone. Its physical meaning depends on the choice of degrees of freedom and on the underlying model. Consequently, specifying an appropriate subsystem structure is a necessary preliminary step in any attempt to reconstruct geometry from quantum information.

The central objective is to investigate whether informational quantities can be mapped consistently onto geometric data. Such a correspondence would need to specify how quantities including entropy, mutual information, and correlation functions relate to candidate notions of distance, dimension, and curvature. In particular, an information-based distance prescription must be supplemented by a construction that satisfies the relevant metric axioms, while a continuum interpretation requires additional assumptions concerning regularity, connectivity, and convergence. These properties are not guaranteed by the existence of quantum correlations alone.

As a first step, we introduce an entanglement-correlation graph \(G=(V,E)\), whose vertices \(v_i\in V\) represent the chosen subsystems and whose edges encode nonzero correlations between them. A possible edge weight is the mutual information,
\[
w_{ij}=I(A_i:A_j), \qquad I(A:B)=S(A)+S(B)-S(A\cup B),
\]
where $S(A)=-\operatorname{Tr}(\rho_A\log\rho_A)$ is the von Neumann entropy of subsystem \(A\). Mutual information is symmetric and non-negative, and it measures total correlations, including both classical and quantum contributions. It therefore provides a useful candidate for quantifying informational connectivity, although it should not be identified with entanglement alone.

Within this construction, stronger correlations may be interpreted as indicating closer relational connections, whereas weaker correlations may correspond to less strongly connected subsystems. This motivates the possibility of defining an effective notion of distance from the correlation structure. Such an interpretation is a modeling hypothesis rather than an established geometric result: mutual information itself is not a metric, and an explicit distance prescription, together with a suitable graph-distance construction, is still required.
At this stage, the graph contains relational information but no prescribed spatial metric or continuum manifold. It can therefore be regarded as a candidate pregeometric structure. The subsequent task is to determine under what conditions, if any, its correlation data can support a consistent metric space and a meaningful continuum limit.

\subsection{Entanglement-Induced Metric Geometry}

Let \(\rho\) be the quantum state of a composite system whose subsystems are denoted by \(A_i\). The mutual information between two subsystems is defined as $I_{ij}=S(\rho_i)+S(\rho_j)-S(\rho_{ij})$, where \(S(\rho)=-\mathrm{Tr}(\rho\ln\rho)\) denotes the von Neumann entropy.
Rather than identifying the logarithmic mutual-information expression (\ref{dist}) directly with a metric distance, we interpret it as an elementary edge length of an entanglement graph. Specifically, we define
\begin{equation}\label{ell}
\ell_{ij}=\ell_0\ln\left(1+\frac{I_0}{I_{ij}}\right),
\qquad I_{ij}>0,
\end{equation}
where \(\ell_0\) is a fundamental length scale and \(I_0\) is a positive reference information scale.
This definition guarantees strictly positive, symmetric edge lengths and ensures that the edge length decreases monotonically with increasing mutual information. The entanglement graph is then equipped with the shortest-path distance
\begin{equation}
d_G(i,j)= \inf_{\gamma:i\rightarrow j} \sum_{(a,b)\in\gamma}\ell_{ab},
\end{equation}
where the infimum is taken over all paths connecting vertices \(i\) and \(j\). For a connected graph with strictly positive edge lengths, \(d_G\) satisfies positivity, identity of indiscernibles, symmetry, and the triangle inequality. Consequently, \((V,d_G)\) constitutes a metric space. Importantly, the triangle inequality follows from the shortest-path construction and does not require strong subadditivity of quantum entropy.

To investigate the continuum limit, we consider a sequence of entanglement graphs whose metric spaces converge, under an appropriate coarse-graining procedure, to a limiting metric space \((X,d)\). The existence of a smooth manifold structure is an additional assumption and is not implied by metric convergence alone. In a regime where the limiting space admits a sufficiently regular Riemannian structure, we introduce Synge's \cite{Sy} world function,
\begin{equation}
\sigma(x,y)=\frac12d^2(x,y).
\end{equation}
Within a sufficiently small geodesically convex neighborhood, the metric tensor can then be reconstructed from the mixed derivatives of the world function:
\begin{equation}\label{metr}
g_{\mu\nu}(x)= -\lim_{y\rightarrow x} \frac{\partial^2\sigma(x,y)} {\partial x^\mu\partial y^\nu}.
\end{equation}
Thus, the emergent metric is not postulated directly from the mutual information but is obtained through a sequence of mathematically distinct constructions: entanglement correlations, graph edge lengths, shortest-path metric, continuum limit, and finally differential-geometric reconstruction.

At microscopic scales, where the limiting geometry may remain fractal or nonsmooth, the metric-tensor description need not be valid. In that regime, the underlying metric-measure structure is regarded as the more fundamental geometric description, while the smooth metric tensor represents an effective macroscopic approximation.

The proposed relationship between entanglement and geometry is conceptually related to holographic approaches, in which geometric quantities are connected to quantum entanglement, as illustrated by the Ryu--Takayanagi relation (\ref{RyuTa}). FEQG explores the possibility of extending this idea from geometric areas to a more general construction of effective distances and metric properties from the correlation structure of a quantum state. However, the reconstruction of a complete metric from entanglement data requires additional assumptions and a well-defined mathematical procedure; it does not follow directly from the Ryu--Takayanagi relation.

A further hypothesis of FEQG is that the resulting geometry may exhibit scale-dependent or fractal properties. Hierarchical entanglement structures, recursive correlations, and coarse-graining procedures provide possible mechanisms through which nontrivial scaling behavior could arise. Nevertheless, hierarchical organization alone does not guarantee fractality. To establish this property, one must demonstrate a suitable scaling law for a specified geometric or graph-theoretic observable.

One possible way to characterize the effective geometric dimension of an entanglement network is through a volume-growth exponent. Let \(N_i(r)\) denote the number of nodes within graph distance \(r\) of a reference node \(i\). For a sufficiently homogeneous network, one may define an effective growth dimension by
\begin{equation}
D_{\mathrm{vol}} =\lim_{r\to\infty} \frac{\log N_i(r)}{\log r},
\end{equation}
provided that the limit exists and is independent of the reference node, at least in an appropriate statistical or coarse-grained sense. More generally, when the scaling behavior changes with distance, a scale-dependent effective dimension may be introduced as
\begin{equation}
D_{\mathrm{vol}}(r)=\frac{d\log N_i(r)}{d\log r}.
\end{equation}
These definitions characterize volume growth on the graph and are related to, but not necessarily identical with, the Hausdorff dimension of a continuum space. In particular, their interpretation depends on the graph metric, the connectivity properties, and the coarse-graining prescription.

If the number of nodes within a graph-distance radius \(r\) follows a power law, $N_i(r)\sim r^{D_{\mathrm{vol}}}$,
the exponent provides an effective measure of the network's volume-growth behavior. A non-integer value may indicate anomalous scaling and can motivate a fractal interpretation. However, such a conclusion requires explicit calculations on the network; it cannot be inferred solely from the presence of entanglement or hierarchical clustering.

Tensor-network constructions, including multiscale entanglement renormalization ansatz (MERA), offer useful examples of hierarchical organization and scale-dependent entanglement. They provide a framework for studying how correlations are distributed across scales and how effective geometric descriptions can be associated with network structure. Nevertheless, the existence of a hierarchical tensor network does not, by itself, establish a non-integer Hausdorff dimension or a unique emergent spacetime geometry. Any such interpretation requires a separate analysis of the relevant graph distances, scaling properties, and continuum observables.

If an appropriate continuum limit exists, the graph distance may provide the starting point for reconstructing an effective metric tensor, as discussed in connection with (\ref{metr}). Such a reconstruction requires more than an increasing density of graph nodes: it also requires suitable regularity, convergence, and local geometric properties. In particular, a sequence of discrete metric spaces does not necessarily converge to a differentiable manifold. Establishing a smooth continuum geometry is therefore a separate mathematical problem within FEQG.

The framework also explores the possibility that the effective geometry may exhibit different scaling regimes. At sufficiently large scales, a continuum approximation could become appropriate, whereas microscopic scales may retain nontrivial graph structure and scale-dependent geometric properties. A possible ultraviolet dimensional flow, $D_{\mathrm{eff}}(\ell)\rightarrow 2$ as $\ell\rightarrow\ell_P$, is one hypothesis that could be investigated within this setting. To establish it, however, one must specify which notion of dimension is being considered and calculate its scale dependence. In particular, a flow of the spectral dimension should be distinguished from a flow of the volume-growth or Hausdorff dimension.

Taken together, these considerations suggest a possible research strategy in which a quantum state determines correlation data, those data define a weighted entanglement graph, and the graph provides a candidate relational structure from which effective geometric properties may be reconstructed. The scaling behavior of this structure can then be investigated using tools from fractal analysis, tensor networks, information geometry, and renormalization-group methods. Whether this procedure yields a consistent fractal continuum geometry and a corresponding gravitational dynamics remains to be demonstrated.

Thus, the central objective of FEQG is to investigate whether quantum information and entanglement can provide a common starting point for the emergence of geometric and dynamical structures. The proposed sequence from quantum correlations to graph geometry, effective dimension, continuum spacetime, and gravitational dynamics should be understood as a research program rather than as an established chain of mathematical implications. Its realization will require explicit constructions, convergence results, and a consistent dynamical formulation.

\subsection{Quantum Mechanics and Gravity Emerging}

We now reach the core of the program, the possibility that both quantum mechanics and
gravity may emerge from a single underlying fractal-entanglement structure.\vspace{0.3cm}

In the FEQG framework, the stochastic behavior of effective trajectories is not assumed to follow automatically from the scale-relativity formalism. Instead, it is proposed that such behavior may arise as a continuum approximation of the underlying entanglement-network dynamics.

Let \(W_{ij}\) denote the transition rate between two network nodes \(i\) and \(j\), and let \(\Delta x_{ij}^{a}\) represent the corresponding effective displacement. In a Markovian approximation, the probability distribution over network nodes evolves according to
\begin{equation}
\frac{dp_i}{dt} = \sum_j\left(W_{ji}p_j-W_{ij}p_i\right).
\end{equation}
If the network admits a suitable continuum limit, with sufficiently local transitions and approximately isotropic statistics, a Fokker-Planck equation may emerge. Its diffusion tensor is determined by the second moment of the transition displacements:
\begin{equation}
B^{ab}(i)=\sum_j W_{ij}\Delta x_{ij}^{a}\Delta x_{ij}^{b}.
\end{equation}
In an isotropic regime, \(B^{ab}\simeq 2D\delta^{ab}\), yielding the effective diffusion coefficient
\begin{equation}
D\simeq \frac{1}{2d}\sum_j W_{ij}\ell_{ij}^{2},
\end{equation}
where $d$ denotes the effective spatial dimension and \(\ell_{ij}\) the corresponding network--link length.
This expression provides a possible connection between microscopic network dynamics and macroscopic stochastic behavior. However, it depends on the assumed transition dynamics and continuum approximation; it is not yet a derived prediction of FEQG.

An important distinction must be made between this stochastic evolution and quantum evolution. A classical diffusion equation does not, by itself, imply the Schr\"odinger equation. The latter additionally requires a coherent dynamical structure and an appropriate effective action scale. Consequently, the identification \(D=\hbar/(2m)\), used in scale-relativity approaches, is not imposed here as an independent microscopic postulate. It should instead be recovered from the effective FEQG dynamics if the proposed emergence mechanism is successful.

The Schr\"odinger equation is therefore regarded, at the present stage, as a candidate effective limit of FEQG rather than as a result already established by the network construction. A complete derivation requires specifying the microscopic transition law, demonstrating the emergence of coherent quantum dynamics, and deriving the effective parameters, including the mass and action scales.

This distinction clarifies the status of the proposal: FEQG provides a possible geometric and informational origin for stochastic dynamics, while the derivation of quantum evolution from that dynamics remains an open problem.

\subsubsection{From Nondifferentiable Geometry to Quantum Dynamics and Emergent Gravity}

Within the FEQG research program, one possible route towards understanding quantum behavior is to investigate the role of nondifferentiable or scale-dependent geometric structures at sufficiently short distances. Below a characteristic scale \(\ell_c\), the assumption of smooth particle trajectories may cease to provide an adequate effective description. In such a regime, classical notions of instantaneous velocity and acceleration may require generalization. This possibility is conceptually related to stochastic and scale-relativistic approaches to quantum mechanics, in which nondifferentiable trajectories and fluctuations play a central role.

In a smooth setting, the forward and backward derivatives of a sufficiently regular trajectory coincide. In a stochastic description one may introduce distinct forward and backward mean velocities, defined schematically by
\begin{equation}
v_{+}=\lim_{dt\to 0^{+}}
\mathbb{E}\left[\frac{X(t+dt)-X(t)}{dt}\right],
\qquad
v_{-}=\lim_{dt\to 0^{+}}
\mathbb{E}\left[\frac{X(t)-X(t-dt)}{dt}\right],
\end{equation}
when these limits exist. Here, the expectation values are understood with respect to the underlying stochastic process. Their distinction reflects the different forward and backward conditional descriptions of the motion. It should not, however, be interpreted as a general consequence of fractal geometry alone: the existence and properties of these velocities depend on the chosen stochastic process and its regularity assumptions.

In scale-relativistic approaches, the two velocity fields can be combined into a complex velocity,
\begin{equation}
V=\frac{v_{+}+v_{-}}{2}-i\,\frac{v_{+}-v_{-}}{2}.
\end{equation}
The real part describes the mean drift, whereas the imaginary part encodes the difference between the forward and backward stochastic descriptions. This complexification provides a useful mathematical representation of the two-valued velocity structure. It is a feature of the adopted formalism, rather than a general mathematical necessity following from nondifferentiability.

A corresponding stochastic description of the trajectory may be written as $dX=b(X,t)\,dt+d\xi$,
where \(b\) denotes a drift field and the fluctuations satisfy, in an isotropic diffusion model,
$\mathbb{E}[d\xi]=0$, $\mathbb{E}[d\xi^a d\xi^b]=2D\,\delta^{ab}\,dt$.
The diffusion coefficient \(D\) characterizes the stochastic process. Such a description is compatible with Brownian scaling and provides a useful model for investigating the effective behavior of nondifferentiable trajectories. Nevertheless, the stochastic process and its diffusion coefficient must be specified independently; neither follows automatically from the existence of a fractal geometric structure.

Within the scale-relativistic formulation, the complex velocity motivates the introduction of a generalized time derivative of the form
\begin{equation}
\widehat{\frac{d}{dt}}=\partial_t+V\cdot\nabla-iD\nabla^2.
\end{equation}
The directional derivative describes transport along the complex velocity field, while the Laplacian term incorporates the contribution associated with the stochastic fluctuations. This operator is a characteristic ingredient of the scale-relativistic construction under its corresponding assumptions. Its use in FEQG should therefore be understood as a candidate effective description whose relation to the underlying entanglement graph remains to be established.

One may then formulate a generalized equation of motion,
\begin{equation}
m\,\widehat{\frac{d}{dt}}V=-\nabla U,
\end{equation}
where \(U\) is an external potential. This equation provides a possible extension of the classical dynamical description to the complex velocity field. However, its application to FEQG requires a derivation of the effective stochastic dynamics and of the generalized derivative from the microscopic graph structure. In the absence of such a derivation, it should be regarded as an adopted effective ansatz rather than a fundamental equation of motion established by the framework.

A further step in the scale-relativistic approach is to introduce a complex action \(S(x,t)\) and relate it to the velocity field through
$ V=\nabla S/m$. The associated wavefunction is defined by $\psi=\exp\left(iS/\hbar\right)$.
This transformation connects the complex action formulation with the wavefunction description of quantum mechanics. Under the assumptions of the scale-relativistic construction, and with the identification $D=\hbar/(2m)$,
the generalized dynamical equation can be related to the Schr\"odinger equation,
\begin{equation}
i\hbar\,\partial_t\psi=-\frac{\hbar^2}{2m}\nabla^2\psi+U\psi.
\end{equation}
This connection illustrates how Schr\"odinger-type dynamics can arise within a particular stochastic, nondifferentiable description of motion. However, the identification of the diffusion coefficient with \(\hbar/(2m)\) is an additional physical input or condition of the construction; it has not been derived from the FEQG entanglement network. Consequently, the emergence of quantum mechanics from the microscopic geometry remains a central conjecture of the framework rather than a completed result.

The second major ingredient of FEQG concerns the possible emergence of geometry from quantum correlations. Given a quantum state \(\rho\) on a collection of subsystems, one can characterize correlations between subsystems \(i\) and \(j\) through their mutual information,
$I_{ij}=S(\rho_i)+S(\rho_j)-S(\rho_{ij})$.
An illustrative prescription for associating a positive link length with these correlations is given by Eq. (\ref{ell}).
A graph distance can then be constructed from shortest paths on the resulting weighted graph. This provides a candidate relational description of geometry based on quantum correlations. It should be emphasized that this construction defines a graph metric under appropriate connectivity assumptions; the emergence of a smooth spacetime metric and its curvature requires additional continuum-limit and regularity conditions.

If the underlying quantum state evolves in time, its correlation structure and the associated graph geometry may also become time-dependent. This suggests a possible connection between quantum information dynamics and effective geometric dynamics. Nevertheless, time dependence of the graph distances does not by itself determine spacetime curvature or gravitational field equations. A complete construction would require a prescription for the continuum geometry, its causal structure, and its dynamical evolution.

In this context, FEQG explores whether gravitational dynamics could emerge as an effective description of the evolution and coarse-graining of the entanglement network. The Einstein tensor \(G_{\mu\nu}\) might then arise in an appropriate macroscopic limit, provided that the effective geometry becomes sufficiently smooth and that the resulting field equations satisfy the required covariance and conservation conditions. Recovering $G_{\mu\nu}=8\pi G\,T_{\mu\nu}$ in the classical regime is therefore an important target of the framework, not a result established by the graph construction alone.

This research direction is conceptually related to holographic approaches, including the Ryu--Takayanagi relation (\ref{RyuTa}), in which geometric quantities are connected to entanglement entropy in specific gravitational settings. It is also related to thermodynamic approaches to gravity, such as Jacobson's derivation of Einstein's equations under suitable assumptions. These connections provide motivation for investigating a common relationship between quantum information, geometry, and gravitational dynamics, but they do not establish the FEQG construction or imply that its field equations follow from those approaches.

Another hypothesis explored by FEQG is that the effective dimension of the underlying geometric structure may vary with scale, potentially approaching a value near two in the ultraviolet regime: $D(\ell)\rightarrow 2$ as $\ell\rightarrow\ell_P$.
Such dimensional flow has been investigated in several approaches to quantum gravity and may have implications for ultraviolet propagation and the behavior of quantum fields. Within FEQG, however, this behavior requires an explicit spectral or geometric calculation on the entanglement graph. Moreover, a reduction of the spectral dimension does not automatically guarantee ultraviolet finiteness or provide a complete regularization of quantum field theory.

Taken together, these ideas suggest a possible sequence of relationships between quantum information, entanglement correlations, emergent geometry, scale-dependent structure, quantum dynamics, and gravitational phenomena. The central hypothesis of FEQG is that these ingredients may be understood within a common framework. Establishing this hypothesis will require deriving the stochastic dynamics and its quantum limit from the microscopic graph, constructing a consistent continuum and relativistic geometry, and obtaining gravitational field equations compatible with covariance and energy-momentum conservation.

Thus, FEQG should presently be regarded as an exploratory research program investigating a possible informational and geometric origin of quantum mechanics and gravity. Its proposed chain of emergence provides a conceptual direction for further work, while the mathematical derivation and physical consistency of its individual links remain open problems.

\subsection{Spectral Dimension and Effective Gravitational Potential}

The reduction of the effective spacetime dimension is treated in FEQG as a hypothesis that must be tested through the spectral properties of the entanglement network. To investigate this behavior, we introduce a weighted graph Laplacian constructed from the network-link lengths.
For a network with positive link lengths \(\ell_{ij}\), we consider the diffusion weights $w_{ij}=\ell_{ij}^{-2}$
for connected pairs of nodes. The corresponding weighted Laplacian is
\begin{equation}\label{lap}
L_{ij}=
\begin{cases}
-w_{ij}, & i\neq j,\\
\sum_{k\neq i}w_{ik}, & i=j.
\end{cases}
\end{equation}
The heat kernel is defined by $K(\tau)=e^{-\tau L}$,
and the averaged return probability for a finite network with \(N\) nodes is
\begin{equation}
P(\tau)=\frac{1}{N}\operatorname{Tr}K(\tau)=\frac{1}{N}\sum_{n=1}^{N}e^{-\tau\lambda_n},
\end{equation}
where \(\lambda_n\) are the eigenvalues of \(L\). The scale-dependent spectral dimension is then defined as
\begin{equation}\label{sdim}
d_s(\tau)=-2\frac{d\ln P(\tau)}{d\ln\tau}.
\end{equation}
A reduction toward two dimensions at short diffusion scales would require an explicit spectral calculation demonstrating a regime in which \(d_s(\tau)\) approaches approximately two. Such a behavior cannot be inferred solely from the fractal character of the network. It must be verified for explicit network realizations and tested for stability as the network size and microscopic parameters are varied.

The gravitational sector requires a separate consistency check. In three spatial dimensions, we consider the fractional Poisson equation
\begin{equation}\label{fPe}
(-\nabla^2)^{\alpha/2}\Phi(\mathbf r) =4\pi G_\alpha\rho(\mathbf r).
\end{equation}
Here, \(G_\alpha\) denotes the coupling constant associated with the fractional operator.
For a point source of mass \(M\), its Fourier-space solution is
\begin{equation}\label{sol}
\widetilde{\Phi}(\mathbf k)
=\frac{4\pi G_\alpha M}{|\mathbf k|^\alpha}.
\end{equation}
For \(0<\alpha<3\), the corresponding position-space potential is
\begin{equation}\label{pot}
\Phi(r)=-4\pi G_\alpha M \frac{\Gamma\left(\frac{3-\alpha}{2}\right)}{2^\alpha\pi^{3/2}\Gamma\left(\frac{\alpha}{2}\right)}
r^{\alpha-3},
\end{equation}
where the sign is chosen so that the potential is attractive.
The conventional Newtonian potential is recovered when \(\alpha=2\), provided \(G_2=G\). This result establishes the mathematical consequence of the assumed fractional Poisson equation, but does not by itself derive that equation from FEQG.

A complete derivation must establish how the continuum gravitational propagator emerges from the entanglement-network Laplacian and whether its spectral scaling leads to the proposed fractional operator. In particular, the fractional order \(\alpha\) should not be identified directly with the spectral dimension without an additional derivation.

Consequently, the ultraviolet dimensional reduction and the modified gravitational potential are treated as testable predictions whose quantitative validity depends on the spectral analysis and continuum-limit construction of the network. Numerical evaluation of the graph spectrum and its Green function is required to determine whether these predictions distinguish FEQG from existing approaches.

\subsection{Observable Predictions and Future Developments}

Within the FEQG framework, a possible phenomenological consequence of scale-dependent or fractal geometry is a modification of the Newtonian gravitational potential. In three-dimensional Euclidean space, the Green's function of the standard Laplacian leads to the familiar behavior \(V(r)\propto 1/r\). A departure from this scaling could arise if the effective spatial geometry or the differential operator governing gravitational propagation differs from its standard form at certain scales. As an illustrative parametrization, one may write
\begin{equation}
V(r)\propto r^{-(1-\varepsilon)},
\end{equation}
where \(\varepsilon\) is a phenomenological parameter that could depend on scale. This expression is intended as a possible effective modification, rather than as a result already derived from the FEQG framework.

A useful mathematical setting for investigating such modifications is the fractional Poisson equation (\ref{fPe}),
where recall that \(\alpha\) denotes the order of the fractional Laplacian and \(G_\alpha\) is a coupling whose physical dimensions depend on \(\alpha\). In three-dimensional Euclidean space, the Green's function of this operator scales, for \(0<\alpha<3\), as $
\Phi(r)\propto r^{\alpha-3}$. Consequently, choosing \(\alpha=2+\varepsilon\) gives $\Phi(r)\propto r^{-(1-\varepsilon)}$ with \(-2<\varepsilon<1\).
The standard Newtonian scaling is recovered for \(\alpha=2\), corresponding to \(\varepsilon=0\).

This calculation illustrates how a modified gravitational potential can arise from a fractional differential operator. However, it does not by itself establish that spacetime fractality leads to this operator, nor does it determine the value or scale dependence of \(\varepsilon\). In particular, identifying the fractional order \(\alpha\) with an effective fractal dimension requires an additional theoretical argument. Establishing such a connection within FEQG would require a well-defined microscopic graph or geometric model, a derivation of its effective continuum operator, and an analysis of the relevant coarse-graining limit. The fractional Poisson equation therefore provides a possible mathematical framework for exploring gravitational corrections, while their derivation and physical relevance remain open questions.

More specifically, we consider a phenomenological extension of the Poisson equation in three spatial dimensions (\ref{fPe})
where the fractional Laplacian is defined by its Fourier-space representation,
$\mathcal F\!\left[(-\nabla^2)^{\alpha/2}\Phi \right](\mathbf k)=|\mathbf k|^\alpha\widetilde{\Phi}(\mathbf k)$.
For a point source of mass \(M\), \(\rho(\mathbf r)=M\delta^{(3)}(\mathbf r)\), the Fourier-space solution is (\ref{sol}).
For \(0<\alpha<3\), the corresponding radial solution is given by (\ref{pot})
and the Newtonian potential is recovered at \(\alpha=2\).
Recall that writing \(\alpha=2+\varepsilon\) gives the scaling $\Phi(r)\propto r^{-(1-\varepsilon)}$ and the corresponding radial acceleration therefore scales as $|g(r)|\propto r^{-(2-\varepsilon)}$.
For positive \(\varepsilon\), the force decreases more slowly with distance than the Newtonian inverse-square law. This behavior may motivate further investigation of the model in the context of galactic dynamics. However, the point-source scaling alone does not establish a dark-matter explanation or reproduce observed rotation curves. Such an interpretation requires a treatment of extended mass distributions, a consistent specification of the coupling \(G_\alpha\), and quantitative comparison with observational data.

One possible phenomenological consequence of a scale-dependent modification of the gravitational potential would be a departure from the Newtonian inverse-square law. For example, if the force law takes the illustrative form
\begin{equation}
F(r)\propto r^{-(2-\varepsilon)},
\end{equation}
with \(0<\varepsilon<1\), the force would decrease more slowly with distance than the Newtonian \(1/r^2\) law. Such a modification could, in principle, contribute to the explanation of anomalous galaxy rotation curves. However, this possibility has not yet been established within FEQG. In particular, it would require a derivation of the scale dependence of \(\varepsilon\), a consistent treatment of extended matter distributions, and quantitative comparisons with astrophysical observations. Conversely, the recovery of Newtonian gravity at laboratory scales would require \(\varepsilon(\ell)\) to become sufficiently small in that regime. Whether a single scale-dependent model can satisfy both astrophysical and terrestrial constraints remains an open question. These considerations therefore identify possible observational regimes rather than demonstrate the phenomenological viability of the model.

A second direction concerns a possible relation between entanglement and spacetime curvature. At a heuristic level, one may investigate whether variations in an appropriately defined entanglement-entropy functional are related to effective geometric quantities. A schematic expression such as
\begin{equation}
R\sim\nabla^2 S_{\mathrm{ent}}
\end{equation}
can be used to represent this motivation, but it should not be interpreted as a derived field equation. A rigorous formulation would require a precise definition of the entropy functional, a specification of the geometric and differential structures involved, and a derivation from an underlying dynamical or variational principle. In particular, the relation between entanglement and curvature need not be local or reducible to a simple scalar equation.

One possible tensorial ansatz for exploring this connection is
\begin{equation}
E_{\mu\nu}=\nabla_\mu\nabla_\nu S_{\mathrm{ent}}-g_{\mu\nu}\Box S_{\mathrm{ent}}.
\end{equation}
This expression is a covariant tensor constructed from an entropy scalar, but it cannot be identified with a conserved information stress-energy tensor without further justification. In particular, its covariant divergence is not generally zero:
\begin{equation}
\nabla^\mu E_{\mu\nu}=R_{\nu}{}^{\lambda}\nabla_\lambda S_{\mathrm{ent}},
\end{equation}
up to the convention adopted for the Riemann curvature tensor. Consequently, incorporating this term into gravitational field equations requires a consistent treatment of the Bianchi identity and of the conservation of the total effective source. A suitable variational formulation or a dynamically specified exchange between sectors would be needed. The expression should therefore be regarded as a phenomenological ansatz, not as an established source of gravitational curvature.

A further possibility concerns the relationship between scale-dependent geometry and quantum decoherence. If quantum dynamics is influenced by fluctuations of an underlying fractal or non-smooth geometric structure, these fluctuations could, in principle, produce corrections to standard quantum evolution. This possibility raises the question of whether some effective decoherence phenomena might have a geometric contribution in addition to the conventional effects associated with environmental interactions. However, such a mechanism has not yet been derived in FEQG, and geometric fluctuations do not necessarily lead to decoherence: their physical effect depends on the underlying dynamics and on whether the evolution of the relevant quantum subsystem is effectively non-unitary.

For illustrative purposes, one may consider a phenomenological modification of the Schr\"odinger equation of the form
\begin{equation}
i\hbar\partial_t\psi=-\frac{\hbar^2}{2m}\nabla^2\psi+U\psi+\hbar\,\delta D(x,\ell)\,\mathcal{O}[\psi],
\end{equation}
where \(\delta D(x,\ell)\) represents a fluctuation of the effective dimension relative to a reference value, and \(\mathcal{O}\) denotes a possible correction operator associated with the underlying geometric structure. This expression is not derived from the FEQG dynamics and should not be regarded as a unique or complete modification of quantum mechanics. Its mathematical consistency would require specifying the operator, its dimensions, its domain, and the conditions under which probability conservation and the appropriate quantum evolution are preserved.

If a consistent dynamical model produces nonzero corrections of this type, possible consequences could include modifications of interference patterns, anomalous propagation, or shifts in energy levels. Atomic spectroscopy and quantum-interference experiments could then provide useful tests, provided that the predicted effects are quantitatively distinguishable from conventional environmental, instrumental, and theoretical corrections. At present, however, no definite magnitude or experimental bound follows from the FEQG framework as formulated here.

Thus, modified gravitational dynamics, entanglement-related geometric effects, and possible corrections to quantum evolution constitute three promising directions for phenomenological investigation. Establishing any of them as a prediction of FEQG will require explicit derivations, well-defined model parameters, and comparisons with existing experimental and observational constraints.

The dimensional flow toward an ultraviolet spectral dimension of \(D=2\), the emergence of Schr\"odinger dynamics, and the proposed entropic modification of the gravitational field equations are not yet derived from a single microscopic dynamics of the entanglement network. In the present formulation, these results should therefore be regarded as conjectured or effective features of the FEQG framework, rather than as established consequences of its network construction.

In particular, the stochastic description used to motivate the Schr\"odinger equation requires a derivation of the relevant transition dynamics and diffusion coefficient from the underlying network. The identification \(D=\hbar/(2m)\) is not assumed to follow from the network geometry alone.

A further consistency condition arises for the proposed entropic tensor,
\begin{equation}
E_{\mu\nu}=\nabla_\mu\nabla_\nu S_{\rm ent} -g_{\mu\nu}\Box S_{\rm ent}.
\end{equation}
Its covariant divergence is not generally zero; it contains a curvature-dependent contribution proportional to \(R_{\nu}{}^\lambda\nabla_\lambda S_{\rm ent}\), with the overall sign depending on the curvature convention. Thus, this tensor cannot be added as an independent source to Einstein's equations without further constraints or additional dynamics.

A consistent gravitational extension must satisfy the contracted Bianchi identity through conservation of the total source. This may require deriving the entropic contribution from a diffeomorphism-invariant action, or specifying dynamical exchange between the entropic and matter sectors. The construction of such a consistent dynamics, together with an explicit network coarse-graining calculation, is left for future work.

Further developing FEQG into a mathematically consistent and predictive framework will require a broader set of tools. Tensor networks may provide a useful language for representing hierarchical correlations and entanglement structures. Fractional and scale-dependent differential operators could be investigated as effective descriptions of nontrivial geometric scaling, provided that their definitions and domains are specified. Quantum information theory offers methods for characterizing correlations, entropy, and quantum channels, while information geometry may help relate statistical structures to candidate notions of distance. Renormalization-group methods could be used to study how effective geometric properties, including the dimension of the underlying structure, vary with scale. Noncommutative geometry is another possible avenue for exploring Planck-scale modifications, although noncommuting coordinates, such as $[x_\mu,x_\nu]\neq 0$,
are not a necessary consequence of the FEQG assumptions developed here.
These mathematical approaches should be regarded as complementary research directions rather than as an already established mathematical backbone of FEQG. Their relevance will depend on whether they can be incorporated into a common formulation and used to derive concrete results.

A further direction is the development of cosmological models that incorporate scale-dependent geometric effects. If dimensional flow can be consistently derived within FEQG, it may affect the evolution of primordial perturbations or lead to scale-dependent features in inflationary observables and the cosmic microwave background (CMB). At present, these possibilities are speculative: identifying an observational signature requires a specific cosmological model, a well-defined perturbation theory, and quantitative predictions that can be compared with data.

More broadly, FEQG explores the hypothesis that spacetime geometry may emerge from the correlation structure of an underlying quantum state, with fractal or scale-dependent properties potentially becoming relevant at short distances. The relations between this geometric structure, quantum dynamics, and gravitational behavior remain conjectural and require explicit derivations. In particular, the framework does not yet establish that fractal motion yields quantum mechanics or that collective entanglement dynamics generate gravitational field equations.

The principal challenge for future work is therefore to turn these conceptual ideas into precise, falsifiable statements. This will require identifying the relevant scales, deriving the associated corrections within a consistent dynamical model, and determining whether those corrections lead to observable effects. Until such calculations are completed, possible cosmological or gravitational signatures should be regarded as motivations for further investigation rather than as predictions of the present framework.

\section{Conclusions}

In this work, the resulting framework may be formulated as a new class of
quantum-gravitational theory, which we have call \emph{Fractal Entanglement Quantum
Gravity} (FEQG). Its central hypothesis is that spacetime is an emergent fractal
geometry generated by the entanglement structure of an underlying quantum information
network, thereby unifying emergent spacetime, quantum mechanics, gravity, holography,
fractal geometry, and renormalization within a single ontological scheme.

The unification achieved by FEQG is conceptual rather than merely formal. Instead of
postulating separate principles for quantum mechanics, spacetime geometry, and
gravitational dynamics, the theory try to derive all three from a single informational
substrate. This provides a coherent ontological picture in which geometry, dynamics,
and quantum behavior are different manifestations of the same underlying structure.

The proposed FEQG framework takes as its fundamental description a triplet \((\mathcal H,\rho,G_E)\), where \(\mathcal H=\bigotimes_i\mathcal H_i\) is the Hilbert space associated with a collection of quantum subsystems, \(\rho\) is their global quantum state, and \(G_E=(V,E)\) is an informational graph whose vertices represent the subsystems and whose edges encode their correlations. No background spatial geometry is assumed at this level. The framework is based on the hypothesis that aspects of spacetime geometry and gravitational dynamics may emerge from the informational structure and evolution of \(\rho\).

For subsystems \(i\) and \(j\), their mutual information is defined as \(I_{ij}=S(\rho_i)+S(\rho_j)-S(\rho_{ij})\). A positive, symmetric link length may be assigned, for example, by $\ell_{ij}=\ell_0\ln\!\left(1+(I_0/I_{ij})\right)$,
for \(I_{ij}>0\), where \(\ell_0\) and \(I_0\) are characteristic scales. The graph distance \(d_G(i,j)\) is then defined as the infimum of the total link length over all paths connecting \(i\) and \(j\). This construction provides a metric on each connected component of the graph, while its interpretation as a continuum spacetime distance requires additional assumptions concerning the continuum limit and regularity of the network.

A central hypothesis of FEQG is that the resulting geometry may exhibit scale-dependent properties, potentially including a reduction of the spectral dimension at short distances. In particular, a flow toward an ultraviolet spectral dimension near two is conjectured, but its establishment requires an explicit network model and a well-defined coarse-graining or diffusion analysis. The Planck length, \(\ell_P=\sqrt{\hbar G/c^3}\), provides a candidate scale at which such effects may become relevant.

The framework also explores the possibility that quantum dynamics can be related to stochastic or nondifferentiable trajectories on an emergent geometry. A stochastic description of the form \(dX^\mu=v^\mu dt+d\xi^\mu\), with vanishing mean fluctuation and a specified covariance, may motivate an effective quantum evolution. However, the stochastic process and its diffusion coefficient--including the identification \(D=\hbar/(2m)\) --are not derived from the informational graph in the present formulation. Accordingly, the emergence of the Schr\"odinger equation is treated as a proposed effective limit rather than as an established consequence of the network construction.

Similarly, the time dependence of the informational correlations may induce an evolving graph geometry and motivates the investigation of an effective gravitational description. Recovering the Einstein field equations in a continuum limit, however, requires a dynamical construction that has not yet been established. In particular, any entropic contribution to the gravitational field equations must be compatible with the contracted Bianchi identity and with conservation of the total energy-momentum source. These consistency conditions remain essential elements of the framework's further development.

Thus, FEQG is presented here as an exploratory framework linking quantum information, graph geometry, scale-dependent structure, and possible effective gravitational dynamics. The proposed connections provide a program for investigation rather than a completed chain of derivations: explicit network calculations, a controlled continuum limit, and a consistent dynamical and conservation structure are required to establish its physical predictions.

A possible effective description of the coupling between matter, entanglement, and scale-dependent geometric effects may be written schematically as
\begin{equation}
G_{\mu\nu} =8\pi G\left(T_{\mu\nu}+\alpha E_{\mu\nu}+\beta F_{\mu\nu}\right),
\end{equation}
where \(T_{\mu\nu}\) is the matter stress-energy tensor, \(E_{\mu\nu}\) denotes a possible contribution associated with the informational structure of the quantum state, and \(F_{\mu\nu}\) represents hypothetical corrections related to scale-dependent or fractal geometry. The coefficients \(\alpha\) and \(\beta\) are coupling parameters whose dimensions and physical interpretation must be specified consistently.

This expression is intended as a phenomenological template rather than as a derived field equation. In particular, the structures of \(E_{\mu\nu}\) and \(F_{\mu\nu}\) are not uniquely determined by the entanglement graph alone. Their construction must also satisfy the geometric and dynamical consistency conditions required for a gravitational theory.

The proposed FEQG framework investigates the possibility that quantum-information correlations define an underlying relational network from which an effective metric description may emerge. Scale-dependent properties of this geometry could, in principle, influence both quantum propagation and gravitational dynamics. However, the emergence of stochastic quantum dynamics, the reduction of the effective spectral dimension, and the recovery of Einstein's equations are distinct problems. Their mutual consistency cannot be assumed without an explicit microscopic model and a controlled continuum limit.

The conjectured ultraviolet flow of the spectral dimension toward \(D\simeq2\) is motivated by dimensional-reduction behavior found in several approaches to quantum gravity. Within FEQG, however, this behavior remains a hypothesis to be tested through an explicit network construction and spectral analysis. Likewise, phenomenological modifications of the gravitational potential require a consistent derivation from the relevant field equation and a quantitative comparison with observational data.

One possible approach is to introduce an effective dimension \(D(x,\ell)\), depending on spacetime position and a characteristic resolution scale \(\ell\), and investigate covariant tensors constructed from this scalar and the curvature. For example, the expression
\begin{equation}
F_{\mu\nu}^{\mathrm{ansatz}}=(D-4)R_{\mu\nu}+\nabla_\mu\nabla_\nu D-g_{\mu\nu}\Box D
\end{equation}
may serve as a phenomenological ansatz, where \(R_{\mu\nu}\) is the Ricci tensor and
\begin{equation}
\Box=g^{\alpha\beta}\nabla_\alpha\nabla_\beta.
\end{equation}
This expression is covariant when \(D\) is a scalar field. If \(D=4\) is constant, it vanishes. These properties make it a possible starting point for investigating scale-dependent corrections. They do not, however, establish that it is the unique or physically appropriate tensor describing fractal geometry. In particular, its covariant divergence is not generally zero. A consistent use of this term in gravitational field equations therefore requires additional dynamics or constraints ensuring conservation of the total source.

Fractional differential operators provide another possible means of representing scale-dependent or nonlocal effects. For example, one may investigate corrections involving a fractional operator acting on curvature,
\begin{equation}
F_{\mu\nu}^{\mathrm{frac}}=\ell_*^{\,2-\sigma}(-\Box)^{\sigma/2}R_{\mu\nu},
\end{equation}
where \(\ell_*\) is a characteristic length and \(\sigma\) is the order of the operator. The precise definition of such an operator in curved spacetime, including its domain, boundary conditions, and tensorial action, must be specified. The expression above is therefore an illustrative proposal, not a uniquely defined FEQG prediction.

Fractional operators can generate nonlocal behavior and power-law modifications in suitable settings. Whether such modifications arise in FEQG, and whether they soften ultraviolet divergences or produce observable effects in gravitational-wave propagation or cosmology, depends on the detailed dynamical model. These consequences cannot be inferred from the presence of a fractional operator alone.

A possible phenomenological hypothesis is that the effective dimension depends on an informational scalar, such as an appropriately defined entanglement-entropy density. For illustration, one might consider
\begin{equation}
D(x,\ell)=4-\eta\,\mathcal S_{\mathrm{ent}}(x,\ell),
\end{equation}
where \(\eta\) is a coupling parameter and \(\mathcal S_{\mathrm{ent}}\) denotes a scalar quantity with a specified normalization and physical dimension.

This relation is not assumed to be a universal identity. Entanglement entropy is generally associated with a subsystem or region, and a local scalar description requires an appropriate prescription. If such a relation is adopted, derivatives of \(D\) can be expressed in terms of derivatives of \(\mathcal S_{\mathrm{ent}}\). This provides a possible phenomenological link between informational structure and geometric corrections, but does not by itself establish that regions of greater entanglement necessarily possess greater curvature.

For example, substituting the illustrative relation into the derivative terms of the preceding ansatz gives
\begin{equation}
\nabla_\mu\nabla_\nu D-g_{\mu\nu}\Box D=-\eta\left(\nabla_\mu\nabla_\nu\mathcal S_{\mathrm{ent}}-g_{\mu\nu}\Box\mathcal S_{\mathrm{ent}}\right),
\end{equation}
provided that \(\eta\) is constant. The resulting tensor must still satisfy the conservation requirements of the full gravitational dynamics. In particular, this relation does not imply that the entanglement contribution is independently divergence-free.

The expressions discussed above illustrate possible ways to parameterize informational and scale-dependent corrections to an effective gravitational description. They do not yet define a unique, self-consistent field theory. A complete formulation must specify the microscopic network, its dynamics, the continuum limit, and the covariant conservation structure of the resulting field equations.

Accordingly, the present framework should be understood as a research program exploring possible connections among quantum information, emergent geometry, scale dependence, and gravitational dynamics. Establishing whether these ingredients yield a consistent theory--and whether they lead to distinctive, testable predictions--requires further mathematical and phenomenological development.

Although the FEQG framework provides a conceptual setting for investigating possible connections between quantum information, emergent geometry, scale dependence, and gravitational dynamics, it does not yet constitute a complete theory of quantum gravity. It is best regarded as an exploratory research program whose proposed physical connections require further mathematical and dynamical development.

A complete formulation must specify its fundamental degrees of freedom, provide well-defined dynamical equations, recover established physics in appropriate limits, and yield quantitative, falsifiable predictions. In the present framework, several of these requirements remain open. In particular, a microscopic evolution law for the universal quantum state \(\rho\) has not yet been established, and the dynamics of the entanglement graph \(G_E\) remain to be specified. A rigorous treatment of scale-dependent or fractal geometry compatible with relativistic covariance is also required.

One possible starting point for the quantum dynamics is an evolution equation for the density matrix,
\begin{equation}
\frac{d\rho}{dt}=\mathcal L[\rho],
\end{equation}
where \(\mathcal L\) is a generator of the evolution. In the special case of closed-system unitary dynamics, this reduces to the von Neumann equation,
\begin{equation}
\frac{d\rho}{dt}=-\frac{i}{\hbar}[H,\rho].
\end{equation}
A possible FEQG implementation would involve a Hamiltonian whose interaction structure depends on the entanglement network, schematically \(H=H(G_E)\). However, the precise form of this Hamiltonian, the dynamical status of the graph, and the manner in which the graph influences the evolution of \(\rho\) have not yet been derived. These expressions therefore identify a possible direction for constructing the dynamics rather than a completed microscopic evolution law.

An alternative approach would be to formulate a path integral over a suitable class of entanglement networks,
\begin{equation}
Z=\sum_{G_E}\mu(G_E)\, e^{iS[G_E]/\hbar},
\end{equation}
where \(S[G_E]\) is an action assigned to each network and \(\mu(G_E)\) denotes an appropriate measure or weighting. Such a formulation would require a precise definition of the admissible graphs, their equivalence classes, the action, and the measure. It would also require a prescription for extracting continuum observables and demonstrating that the resulting theory possesses the desired physical limits. The path-integral expression is thus a possible formal direction, not yet a defined quantization of FEQG.

Further open problems include the construction of a consistent relativistic quantum field theory, the treatment of black-hole entropy and evaporation, the development of cosmological solutions, and the identification of observational signatures that distinguish FEQG from existing approaches. The proposed dimensional flow toward a spectral dimension near two, the emergence of quantum dynamics from the network, and the modification of gravitational interactions must each be supported by explicit calculations within a specified model.

The main challenges for future work can therefore be summarized as follows: (1) Microscopic dynamics: specify the fundamental variables and derive an evolution law for the quantum state and entanglement network; (2) Continuum and relativistic limit: establish a controlled continuum description and determine under what conditions relativistic covariance and general relativity are recovered; (3) Quantum dynamics: derive, rather than assume, the effective quantum evolution and its relevant parameters from the network; (4) Scale-dependent geometry:define the spectral or effective dimension through explicit network calculations and a well-defined coarse-graining procedure. (5) Gravitational consistency: construct field equations whose total source satisfies the conservation conditions required by the contracted Bianchi identity. (6) Phenomenology: obtain quantitative predictions and compare them with observational or experimental constraints.

These requirements define a roadmap for the further development of FEQG. Its central hypothesis is that informational correlations may provide an underlying structure from which aspects of geometry and gravitational dynamics emerge. Whether this hypothesis can be developed into a mathematically consistent and empirically distinguishable theory remains an open question.

\section{Final remarks}

The Fractal Entanglement Quantum Gravity (FEQG) framework is motivated by the possibility that effective spacetime geometry and gravitational interactions originate from the informational structure of an underlying quantum system. Although this proposal draws on established concepts from quantum information theory, holography, tensor networks, fractal geometry, and scale relativity, its purpose is not merely to combine these approaches, but to investigate whether they can be connected through a common mathematical structure.

The fundamental object of the framework is a quantum state described by a density operator \(\rho\) on a Hilbert space decomposed into subsystems, $\mathcal H=\bigotimes_{i=1}^{N}\mathcal H_i$. The correlations between subsystems are characterized by their mutual information, $I_{ij}=S(\rho_i)+S(\rho_j)-S(\rho_{ij})$, where \(S(\rho)=-\operatorname{Tr}(\rho\ln\rho)\) denotes the von Neumann entropy.

These correlations define an informational network whose nodes represent quantum subsystems and whose links encode their correlations. An effective link length can be introduced as $\ell_{ij}=\ell_0\ln\left(1+I_0/I_{ij}\right)$,
for \(I_{ij}>0\). The graph distance is subsequently defined by minimizing the total link length over all paths connecting two nodes.

This construction provides a discrete metric structure associated with the informational organization of the quantum state. It constitutes a proposed geometric prescription within FEQG rather than a uniquely determined consequence of quantum information theory.
To investigate the propagation and spectral properties of this emergent geometry, we introduce a weighted graph Laplacian given by Eq. (\ref{lap}), where, as a minimal model, the weights are chosen as $w_{ij}=\ell_{ij}^{-2}$.
This operator provides a common mathematical object for studying diffusion, spectral dimension, and effective propagation on the entanglement network.

The corresponding heat kernel is $K(\tau)=e^{-\tau L}$, and its normalized return probability is
$P(\tau)=\operatorname{Tr}(e^{-\tau L})/N$. The scale-dependent spectral dimension is then defined as Eq. (\ref{sdim}).
Within FEQG, the possible emergence of scale-dependent dimensionality is investigated through this spectral quantity rather than assumed a priori.

The central research hypothesis is that the informational network and its associated spectral operator may provide a common foundation for effective geometric and physical descriptions. In particular, the spectral properties of \(L\) may determine the behavior of propagation processes and the structure of continuum effective operators. Whether these structures can reproduce gravitational dynamics and quantum evolution remains to be established through explicit derivations. The proposed framework therefore distinguishes three levels of development: (1) Microscopic informational structure: the quantum state and its correlation network; (2) Emergent geometric structure: the graph metric and spectral properties of the associated Laplacian; (3) Effective physical dynamics: the possible emergence of continuum propagation, quantum evolution, and gravitational behavior.

The original contribution sought by FEQG is the investigation of the relationships between these three levels within a unified network-based construction. In particular, the framework proposes that geometric and dynamical properties should be derived from the same underlying informational structure, rather than introduced as independent assumptions.

At the present stage, this proposal should be regarded as a mathematical research framework rather than a completed theory of quantum gravity. The derivation of a consistent microscopic evolution law, the recovery of the Schr\"odinger and Einstein equations, and the demonstration of a distinctive gravitational prediction remain open problems.

The predictive content of FEQG will therefore depend on the quantitative evaluation of explicitly constructed quantum networks, including their spectral dimensions, Green functions, and continuum limits. These calculations are necessary to determine whether the framework produces physical results that differ from those of existing approaches.

A possible conceptual connection exists between the FEQG framework and Bianconi's proposal \cite{GB}, although the two approaches should not be regarded as equivalent at this stage. Bianconi formulates gravity in terms of an entropic action principle, in which a relative-entropy functional involving matter and geometry is used to obtain gravitational field equations. FEQG, by contrast, explores the possibility that the relevant geometric and entropic structures arise from the quantum state and its correlation network. This suggests a potential relationship between the two frameworks, but no mathematical equivalence has been established.

In particular, FEQG does not yet provide a derivation showing that coarse-graining its underlying entanglement graph yields Bianconi's entropic action. Such a result would require a well-defined microscopic dynamics, a coarse-graining prescription, and an effective action whose continuum limit can be calculated. The conjecture $S_{\mathrm{eff}}[g]{=}S_{\mathrm{Bianconi}}[g]$
may therefore be regarded as a possible direction for future investigation, rather than a result of the present framework. Establishing this relation would clarify whether Bianconi's theory can be recovered as an effective macroscopic description of FEQG. At present, however, this remains an open question. Likewise, the recovery of standard quantum mechanics, Einstein gravity, and quantum field theory on curved spacetime should be understood as goals of the framework, not as limits that have already been rigorously derived.

Recent studies further motivate the exploration of connections between quantum information, emergent geometry, and gravitational dynamics. In particular, the derivation of semiclassical Einstein equations from quantum relative entropy highlights how information-theoretic constraints may encode gravitational relations \cite{DM}, while recent work on emergent holographic spacetime emphasizes the role of quantum information in the emergence of geometric structure \cite{Tak}. Cosmological approaches based on uncertainty relations also suggest possible links between quantum principles and the late-time acceleration of the Universe \cite{Kou}. These developments provide complementary motivations for FEQG, which investigates whether the correlation structure of an underlying quantum network can serve as a common pregeometric framework. The connections are suggestive rather than conclusive: establishing how such ideas arise within FEQG requires an explicit dynamical formulation and further derivation.

More generally, FEQG investigates the possibility that spacetime geometry is not fundamental, but emerges from the correlation structure of an underlying quantum state. In this picture, an entanglement graph provides a candidate microscopic description from which geometric and dynamical properties may emerge. The specific mechanisms connecting this graph to a continuum spacetime, quantum dynamics, and gravitational field equations remain to be established.

The central motivation of FEQG is thus to explore a common informational origin for geometry, quantum phenomena, and gravitation. Entanglement may provide the relational structure from which geometry is constructed, while scale-dependent or fractal features may characterize its microscopic regime. Whether these ingredients can be assembled into a mathematically consistent and empirically testable theory remains an open problem. FEQG should therefore be viewed as an exploratory research program whose central conjectures require further mathematical development and explicit physical tests.

Despite its proposed unifying structure, FEQG remains at an exploratory stage. Several mathematical and physical aspects require further development. First, the microscopic dynamics governing the evolution of the quantum state and the associated entanglement network must be specified consistently. Second, the continuum limit of the weighted graph Laplacian must be established, including the conditions under which a smooth or fractal effective geometry emerges. Third, the proposed spectral dimensional reduction must be demonstrated through explicit numerical calculations. Fourth, the effective gravitational propagator and the corresponding gravitational potential must be derived from the network structure rather than introduced independently. Finally, the emergence of coherent quantum dynamics and the recovery of the Schr\"odinger equation require a derivation beyond the adoption of existing scale-relativity assumptions.

Future work will focus on constructing explicit quantum-network models, calculating their spectral properties, and investigating the relationship between their Green functions and effective gravitational interactions. These steps will provide a basis for determining whether FEQG leads to distinctive and experimentally testable predictions.

\end{document}